# New Indivisible Geoscience Paradigm

J. Marvin Herndon
Transdyne Corporation
11044 Red Rock Drive
San Diego, CA 92131 USA
mherndon@san.rr.com

**Abstract:** Earth's interior, I posit, is like one of the rare, oxygen-starved "enstatite chondrite" meteorites (and unlike a more-oxidized "ordinary chondrite" as has been believed for seventy years). Laboratory-analyzed enstatite-chondrite samples are comparable to having-in-hand impossible-to-gather deep-Earth samples. Enstatite-chondrite formation in oxygen-starved conditions caused oxygen-loving elements to occur, in part, as non-oxides in their iron-alloy. Observations, consistent with solar abundance and behavior of chemical elements, lead me to a new interpretation of: (1) Earth's early formation as a Jupiter-like gas-giant, (2) its decompression-powered surface geology, (3) Earth's internal composition, and (4) a natural, planetocentric nuclear-fission reactor as source of both the geomagnetic field and energy channeled to surface "hot-spots". I present a unified vision of Earth formation and concomitant dynamics that explains in a logical and causally related way: (1) fluid Earth-core formation without whole-planet melting, and (2) the myriad measurements and observations, previously attributed to "plate tectonics", but without necessitating mantle convection.

## Background

Wiechert (1897) observed that Earth's mean density (5.5 g/cm$^3$) is too great for the planet to be composed solely of non-metallic rocks. Our planet has a central dense core similar to meteorites composed of nickel-iron metal on display in natural history museums, he postulated. The Earth's core was discovered by Oldham (1906). Earthquake waves travel faster with depth but at 2900km from the surface, their speed abruptly decreases. They enter a different material, postulated by Oldham to be the core. Based on interpretation of many earthquake waves, the core was correctly inferred to be liquid.

Earth's layered structure, by the early 1930s, was asserted to consist of a fluid core, 3500km in radius, surrounded by a uniform 2900km thick solid rock shell, called the mantle, topped by a 10-50km thick crust. Analysis of records of a surprisingly large earthquake near New Zealand led Lehmann (1936) to discover the Earth's almost-Moon-sized inner core, correctly estimated to be 1200km in radius (Figure 1). Although decades later additional data led to refinement of Earth's interior composition, by 1940 this view of Earth's innards became the foundation for most textbook Earth science.

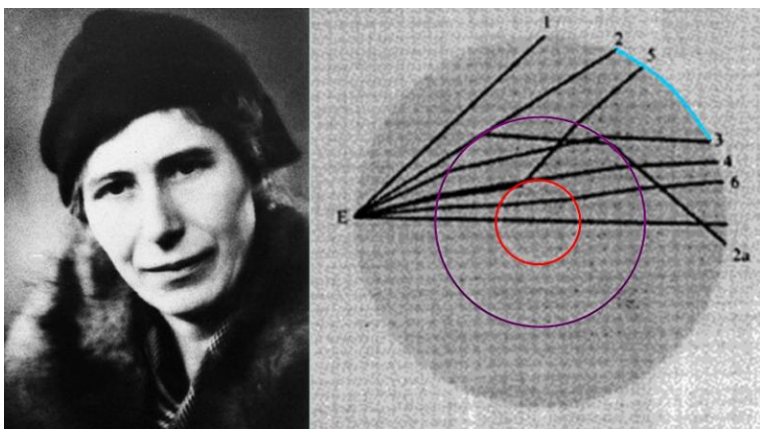

**Figure 1. Inge Lehmann (1888-1993) and her discovery-diagram of Earth's inner core.** For clarity, I traced the circles that bound the inner (red) and fluid (purple) core, and the region where earthquakes were claimed to be undetectable (blue). Ray #5 is reflected into that zone from the inner core she envisioned existed. The only difference between modern representations is that the rays are now known to be slightly curved rather than being straight lines. From Lehmann (1936).

The physical structure of Earth's interior is deducible from seismic observations, but the chemical composition of these inaccessible regions must be inferred from meteorites. Abundant data indicate that Earth, our Moon, meteorites and other Solar System bodies formed ca. 4,500 million years ago from "primordial" matter of uniform composition (Dalrymple 1991). That primordial composition is seen today in the outer part (photosphere) of the Sun and in the non-gaseous elements of chondrite-meteorites. The importance of chondrites is that their non-gaseous elements never appreciably separated from one another as they did in other, more changed meteorites. Consequently, chondrites are appropriately accepted to resemble Earth's bulk chemical composition. But, there are different types of chondrites characterized by strikingly different mineral composition. In my view the incorrect choice of chondrite led to an erroneous assumption of Earth's internal mineral composition. This has long confused geophysicists' ideas about Earth's origin and dynamics.

The three groups of chondrite-meteorites, *carbonaceous*, *enstatite*, and *ordinary* differ markedly in oxygen content and, hence, in mineral composition. Although Lehmann (1936) correctly deduced the presence of Earth's inner core, Birch (1940), on the basis of geophysical understanding in the late '30s and early '40s, incorrectly postulated its composition. The Earth was assumed to be derived from ordinary chondritic materials: silicate-rock minerals and nickel invariably combined with metallic iron. But the solar abundance of elements heavier than nickel and iron, even together, are insufficient to form such a massive inner core. These considerations led Birch (1940) to postulate, analogous to an ice-cube that freezes in a glass of ice-water, that the inner core is composed of nickel-iron metal in the process of freezing from the liquid nickel-iron core.

Nearly four decades after Birch's ideas were entrenched in the literature of geophysics, I investigated enstatite chondrites. Combined with two 1960s discoveries, my studies generated a fundamentally different view of Earth's inner core composition. What was the news? (1) Elemental silicon occurs in the metallic iron of enstatite chondrites (Ringwood 1961), and (2) perryite, nickel-silicide, $Ni_2Si$, is present in many enstatite meteorites (Ramdohr 1964; Reed 1968). I realized that in Earth's core, silicon in chemical combination with nickel would have settled by gravity to the center and, in principle, formed a mass virtually identical to the relative mass deduced for the inner core. When my nickel-silicide-inner-core concept was accepted for publication (Herndon, 1979), Inge Lehmann commented: "I admire the precision of your reasoning based upon available information, and I congratulate you on the highly important result you have obtained." At the time I no idea just how "highly important" the result would become, although I saw clearly how to show conclusively, if the Earth is like a chondrite meteorite as widely believed, it is like an enstatite, not an ordinary, chondrite (Figure 2) and that the relative masses of inner parts of Earth, derived from seismic data, exactly match the corresponding, chemically-identified, relative masses of enstatite-chondrite-components, observed by microscopic examination (Figure 3, Table 1).

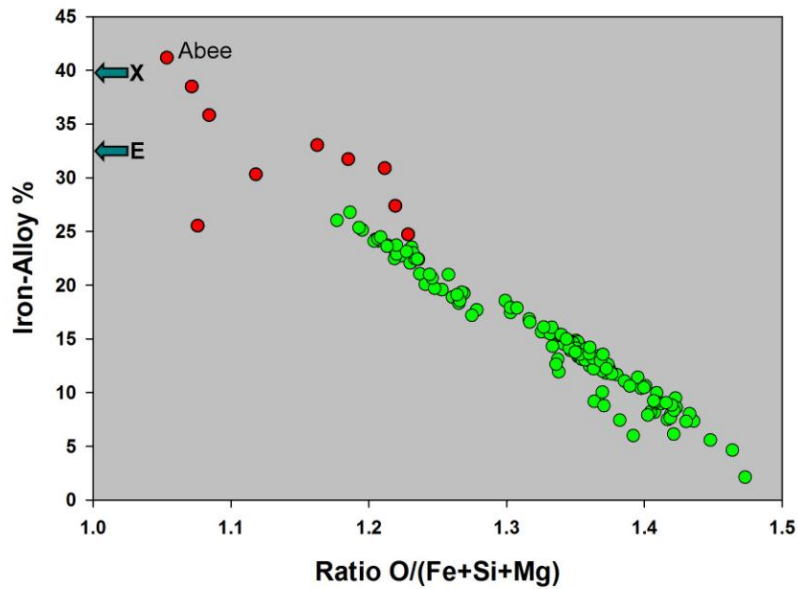

**Figure 2. Evidence that Earth is like an enstatite-chondrite.** The percent alloy (iron metal plus iron sulfide) of 157 ordinary (green)- and 9 enstatite-chondrite-meteorites (red) plotted against oxygen content. The core percent of the whole-Earth, "arrow E", and of (core-plus-lower mantle), "arrow X", shows that Earth is more an Abee-type enstatite-chondrite and not an ordinary-chondrite.

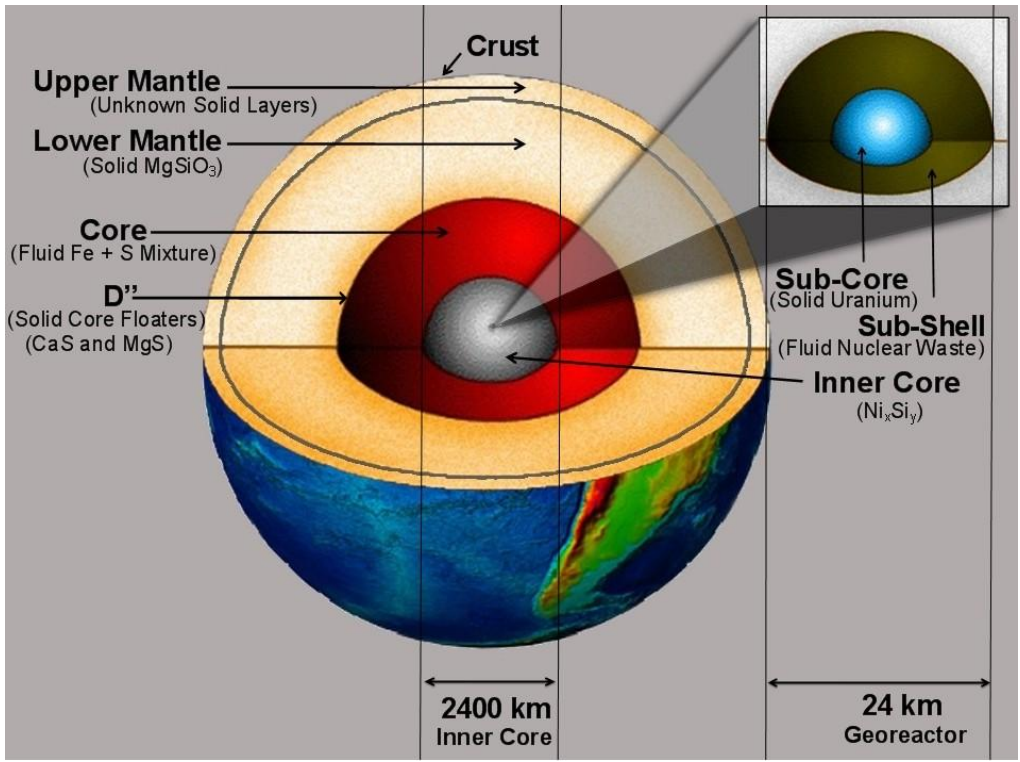

**Figure 3. Earth's composition derived from the Abee enstatite chondrite.** Chemical compositions of the major parts of the Earth, inferred from the Abee enstatite chondrite (see Table 1), and (inset) nuclear-fission georeactor at Earth's center. The upper mantle, above the lower mantle, has seismically-resolved layers whose chemical composition is un known. The georeactor at the center is one ten-millionth the mass of Earth's fluid core. The georeactor sub-shell, I posit, is a liquid or a slurry and is situated between the nuclear-fission heat source and inner-core heat sink, assuring stable convection, necessary for sustained geomagnetic field production by convection-driven dynamo action in the georeactor sub-shell (Herndon 1996, 2007, 2009).

**Table 1. Fundamental mass ratio comparison between the endo-Earth (lower mantle plus core) and the Abee enstatite chondrite. Above a depth of 660 km seismic data indicate layers suggestive of veneer, possibly formed by the late addition of more oxidized chondrite and cometary matter, whose compositions cannot be specified at this time.**

| Fundamental Earth Ratio | Earth Ratio Value | Abee Ratio Value |
|---|---|---|
| lower mantle mass to total core mass | 1.49 | 1.43 |
| inner core mass to total core mass | 0.052 | theoretical 0.052 if $Ni_3Si$ 0.057 if $Ni_2Si$ |
| inner core mass to lower mantle + total core mass | 0.021 | 0.021 |
| D″ mass to total core mass | 0.09‡ | 0.11* |
| ULVZ† of D″ CaS mass to total core mass | 0.012⌡ | 0.012* |

\* = avg. of Abee, Indarch, and Adhi-Kot enstatite chondrites
D″ is the "seismically rough" region between the fluid core and lower mantle
† ULVZ is the "Ultra Low Velocity Zone" of D″
‡ calculated assuming average thickness of 200 km
⌡ calculated assuming average thickness of 28 km
data from Dziewonski & Anderson (1981); Keil (1968); Kennet et al. (1995)

## Jupiter-like Earth Formation

Since the first hypothesis about the origin of the Sun and the planets was advanced in the latter half of the 18$^{th}$ Century by Immanuel Kant and modified later by Pierre-Simon de Laplace, various ideas have been put forward. Generally, concepts of planetary formation fall into one of two categories that involve either (1) condensation at high-pressures, hundreds to thousands of atmospheres; or (2) condensation at very low-pressures.

Throughout the formative period of plate tectonics, from 1963 to the present, the scientific community wrongly concurred Earth formed from primordial matter that condensed at low-pressure, one ten-thousandth of an atmosphere (Cameron 1963). The "planetesimal hypothesis" was accepted as the "standard model of solar system formation". But such low-pressure condensation would lead to terrestrial planets with insufficiently massive cores, as iron would form iron-oxide and not remain as metal (Herndon 2006b).

Thermodynamic considerations led Eucken (1944) to conceive of Earth formation from within a giant, gaseous protoplanet when molten-iron rained out to form the core and when then followed by the condensation of the silicate-rock mantle. By similar, extended calculations I verified Eucken's results and deduced that oxygen-starved, highly-reduced matter characteristic of enstatite chondrites and, by inference, also the Earth's interior condensed at high temperatures and high pressures from primordial Solar System gas under circumstances that isolated the condensate from further reaction with the gas at low temperatures (Herndon 2006b; Herndon & Suess 1976).

The gaseous portion of primordial Solar System matter, as is the Sun's photosphere today, was about 300 times as massive as all of its rock-plus-metal forming elements. Earth's complete condensation formed a gas-giant planet virtually identical in mass to Jupiter. Giant gaseous planets of Jupiter size are observed in extrasolar systems closer to their star than Earth is to the Sun (Seager & Deming 2010). So, what happened with the weight of their gases?

## Removing the Gases

Of the eight planets in the Solar System, the outer four (Jupiter, Saturn, Uranus, and Neptune) are gas-giants, whereas the inner four are rocky (Mercury, Venus, Earth, and Mars), without primary atmospheres. But the inner planets originated as gas-giants and their massive, gaseous envelopes were lost. How?

A brief period of violent activity, the T-Tauri phase, occurs during the early stages of star formation with grand eruptions and super-intense "solar-wind". The Hubble Space Telescope image of an erupting binary T-Tauri star is seen here in Figure 4. The white crescent shows the leading edge of the plume from a five-years earlier observation. The plume edge moved 130AU, a distance 130 times that from the Sun to Earth, in just five years. A T-Tauri outburst by our young Sun, I posit, stripped gas from the inner four planets (Figure 5). A rocky Earth, compressed by the weight of primordial gases, remained.

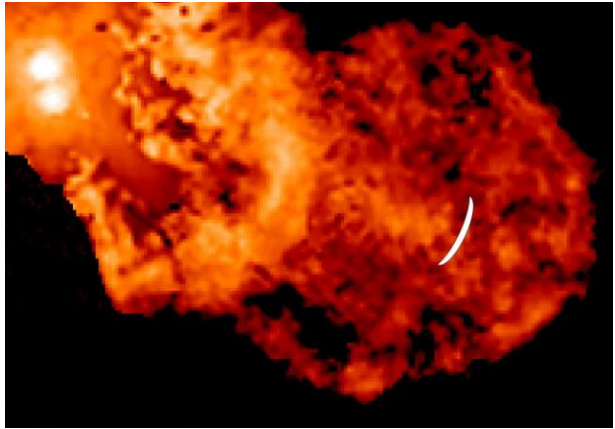

**Figure 4. Outburst from a T-Tauri phase binary star in 2000.** The white crescent label shows the position of the leading edge of that plume in 1995, indicating a leading-edge advance of 130 AU in five years. T-Tauri eruptions are observed in newly formed stars. Such eruptions in the pre-Hadean, I submit, stripped the primordial gases from the inner four planets of our Solar System. [Hubble Space Telescope image of XZ-Tauri (2000).]

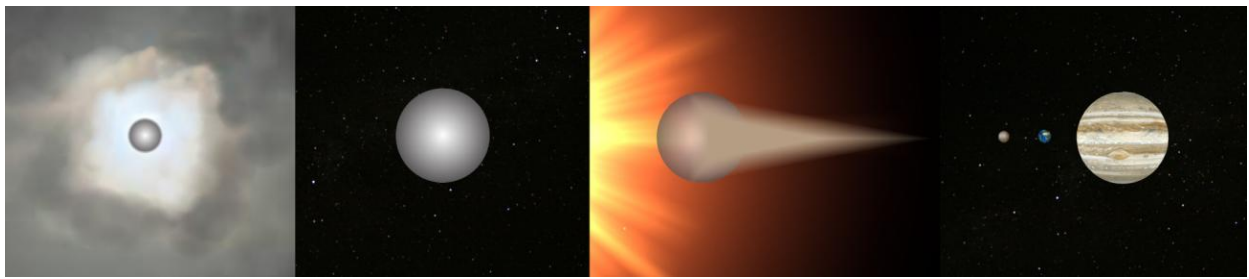

**Figure 5. Whole-Earth decompression dynamics formation of Earth.** From left to right, same scale: 1) Earth condensed at the center of its giant gaseous protoplanet; 2) Earth, a fully condensed a gas-giant; 3) Earth's primordial gases stripped away by the Sun's T-Tauri super-intense solar-wind; 4) Earth at the onset of the Hadean eon, compressed to 64% of present diameter; 5) Earth at Holocene; 6) Jupiter for size comparison.

## Resolution of Geophysical Cognitive Dissonance

The apparent "fit" of transoceanic continental coastlines (South America with West Africa; North America with West Europe), the matching of *Mesosaurus* fossils in Brazil and those in Ghana, match of sediments including coal field strata deposited in the Carboniferous in both Europe and North America led Wegener (1912) to conclude that these continents were joined 330±30mya in the super-continent, Pangaea. The super-continent broke apart and its components drifted for 300 million years through the surrounding ocean to their present locations.

Magnetic striations and great stratified sedimentary layers symmetric on either side of the oceanic ridges, both in the mid-Atlantic and off the North Western coast of North America, led to plate tectonics theory. "Plate tectonics" seemed to explain "continental drift" and is based upon observations that molten basalt-rock extrudes at oceanic ridges, producing symmetric seafloor spreading across the ocean basin. In geologically relative short times, fewer than 170 million years, the seafloor "subducts". Lithospheric solid seafloor is assumed to plunge into the mantle and, as a great conveyor belt loop, to circulate through the mantle, and to re-extrude again at the oceanic ridge zones. Plate tectonics theory depends critically upon the unproven assumption that mantle convection exists and is the motive-force responsible for continent displacement.

Although I entirely accept the young date for the age of the ocean's oldest rocks (170my), the observations of matching transoceanic coastlines with complementary fossil commonalities, conformably overlying sediments separated by immense distances, basaltic seafloor formation at oceanic ridges, seafloor spreading that generates great complementary symmetric magnetic striations, I dispute the existence of mantle convection (Appendix I).

Unlike Wegener's Pangaea, surrounded by a great ocean basin, Hilgenberg (1933) envisioned one continent without ocean basins on a globe smaller than Earth's present diameter that subsequently expanded in a process that fragmented and separated continental masses and formed interstitial ocean basins. Hilgenberg's fundamentally important concept provided the basis for "Earth expansion theory" (Carey 1976). But Earth expansion theory as formulated was unable to explain the reason for Earth's initially smaller size or to provide an explanation for seafloor features. Moreover: "If expansion on the postulated scale occurred at all, a completely unknown energy source must be found" (Scheidegger 1982). I disclose that unknown energy source and resolve geophysical cognitive dissonance.

I unify "plate tectonics" and "Earth expansion" into a new, geological paradigm; I name it *whole-Earth decompression dynamics* (wEdd). Ocean floor topography, distribution of global tectonic activity, and the myriad measurements and observations that underlie Wegener's and Hilgenberg's visions, and their modern expressions, I unite, without mantle convection, in "whole-Earth decompression dynamics". The driving-energy source and surface geodynamics are direct consequences of our planet's early formation as a Jupiter-like gas giant (Herndon 2005, 2006, 2008, 2010).

**Whole-Earth Decompression Dynamics**

The weight of 300 Earth-masses of primordial gases gravitationally compressed the non-gases: the original rock-plus-alloy kernel that became Earth to some 64% of its present radius, sufficient compression for solid continental-rock crust to cover the entire planet, as Hilgenberg envisioned.

As the Sun ignited, its violent T-Tauri solar-wind stripped Earth of its Jupiter-like gas envelope, marking the beginning of the Hadean eon (Figure 5). When internal pressures built up later, sufficient to crack the rigid crust, the Earth began to decompress by expansion (Figure 6). Gravitational energy of compression that had been stored during Earth's gas-giant stage powers Earth dynamics.

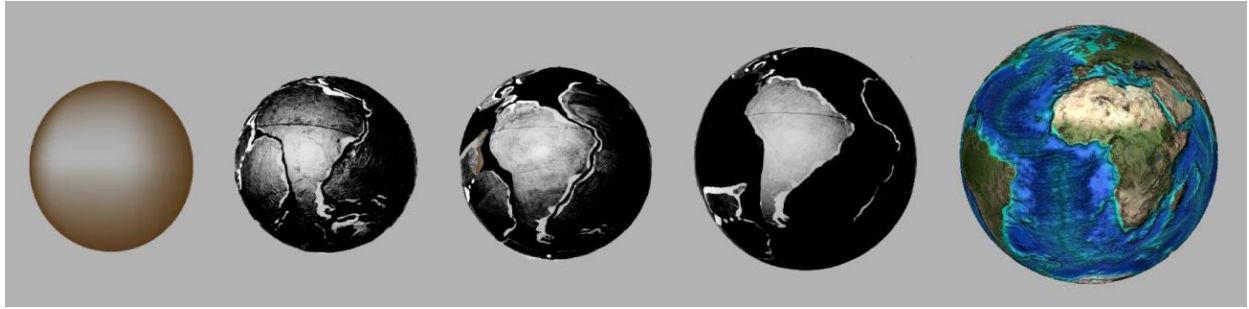

**Figure 6. Decompression of Earth (wEdd) from Hadean to present.** From left to right, same scale: 1) Ottland, 64% of present Earth diameter, fully covered with continental-rock crust; 2), 3), & 4) Formation of primary and secondary decompression cracks that progressively fractured Ottland to open ocean basins. Timescale not precisely established; 5) Holocene Earth.

Whole-Earth decompression produced *primary* cracks in rigid crust. Identified as the oceanic ridge system, they persist. Here ends the similarity with Earth expansion theory. Whole-Earth decompression dynamics identifies *secondary* decompression cracks. Along continent margins they are identifiable as submarine trenches.

*Primary* decompression cracks, with their underlying heat sources, extrude basalt-rock, whereas *secondary* decompression cracks lack heat sources. They became ultimate repositories for extruded basalt-rock. Basalt-rock, extruded from mid-oceanic ridges, traverses the ocean floor by gravitational creep. Ultimately, in a process of "subduction" that lacks any mantle convection, seafloor basalt, with its carbonate sediment, fills in *secondary* decompression cracks (Figure 7). Seismically imaged "down-plunging slabs", I submit, are in-filled *secondary* decompression cracks.

Whole-Earth decompression dynamics extends "plate tectonic" concepts as it is responsible for Earth's well-documented features. Partially in-filled *secondary* decompression cracks uniquely explain oceanic troughs, inexplicable by "plate tectonics". And, compression heating at the base of the rigid crust is a direct consequence of mantle decompression (Herndon 2006a).

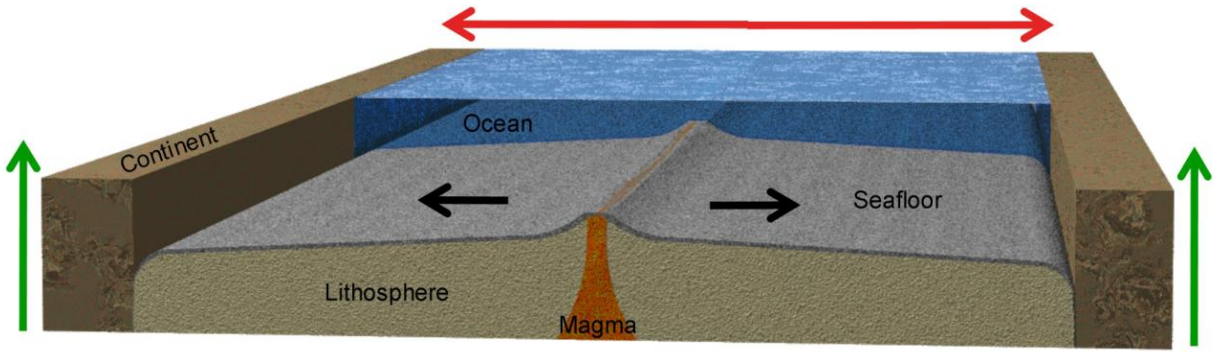

**Figure 7. Ocean floor topography formed by whole-Earth decompression dynamics.** As the Earth decompresses by wEdd, primary cracks with underlying heat sources and secondary cracks that lack heat sources are generated to accommodate decompression. Basalt-rock, extruded at mid-ocean ridges, creeps across the ocean basins (black arrows) to fill in *secondary* decompression cracks often located at continental margins. As the Earth decompresses by wEdd (green arrows), crustal extension (red arrows) forms *secondary* decompression cracks into which seafloor falls; the dynamic processes require no invocation of mantle convection, but their outcomes resemble "subducted tectonic plates".

At onset of the Hadean eon, one ancient supercontinent existed: the 100% closed contiguous shell of continental-rock I name *Ottland*, in honor of Ott Christoph Hilgenberg, who first conceived its existence. The successive fragmentation of Ottland to form new Earth-surface area to accommodate decompression-increased planetary volume, even with competitive interactions between fragments, such as the India-Asia collision, bears little resemblance to the popular, but hypothetical, breakup of Wegener's Pangaea, wherein the continents are assumed free to wander, breaking up and re-aggregating.

### Deep-Earth Composition

Enstatite chondrites, e.g., the Abee meteorite collected in Western Ontario, Canada, in 1952, and by implication, the deep interior of Earth, formed under such severe oxygen-limiting conditions that even elements with a high affinity for oxygen, including calcium, magnesium, and silicon, could not be fully accommodated as oxides; a portion of those elements occurs in the iron-alloy.

Oxygen-high-affinity elements, incompatible in iron alloy, readily precipitate with cooling. In the Earth's core, silicon precipitated in combination with nickel. The nickel-silicide sank by gravity and formed the inner core. Magnesium and calcium that react with sulfur at a high temperature formed sulfides, MgS and CaS, that floated to form the thin, "seismically rough" layers between Earth's fluid core and its enstatite-composition-silicate-rock lower mantle (Figure 3, Table 1).

Silicon, magnesium, and calcium occur partially in the Earth's iron-alloy core because of limited oxygen-availability; otherwise, they would have ended up totally as oxides in Earth's crust or mantle. Uranium, an oxygen-reactive trace element, occurs exclusively in the iron-alloy portion of the Abee enstatite chondrite, and by implication, in the Earth's core. The inferred occurrence of uranium in the Earth's core led me to disclose a powerful, but unanticipated deep-Earth energy source: self-sustaining nuclear fission chain reactions (Herndon 1993, 1994, 2003, 2007).

## Deep-Earth Nuclear Reactor

Uranium, under hot, dense conditions in the Earth's core, is expected to come out of solution as the high-temperature, high-density precipitate uranium sulfide (or even as a metal) and sink to Earth's center by gravity, directly or in stages. The gravity-accumulated uranium, a sphere, estimated to be 20-40km in diameter, beneath the 2400km diameter inner core at the planet's center (Figure 3), will act as a natural nuclear fission reactor (Herndon, 1993, 1994, 1996, 2007). The "georeactor", I posit, consists of a fissioning, nuclear reactor sub-core surrounded by a sub-shell of radioactive waste (fission and decay products) likely to be a fluid or slurry. The dynamic interaction of nuclear heat-production and uranium (or uranium sulfide) settling-out from the fluid sub-shell assures steady energy output (Herndon 2009).

The georeactor, I submit, energizes and produces the Earth's magnetic field, and powers "hot-spot" volcanism. Geological observations of helium isotopic ratios ($^3$He/$^4$He) provide compelling evidence for the nuclear georeactor at Earth's center (Herndon, 2003, 2010, Hollenbach & Herndon 2001). Georeactor-powered volcanism, identified by high $^3$He/$^4$He ratios, formed the Hawaiian Islands and Iceland, and currently occurs beneath continental masses at Yellowstone (U. S. A.), and the Afar triangle in the East African Rift System. Georeactor-produced heat was also involved in the massive flood basalts of the Deccan Traps of India (65mya) and the Siberian Traps (250mya).

## Earth's Magnetic Field and its Reversals

Earth has a centrally-generated magnetic field that directs compass needles and diverts the charged particles of the solar-wind around and past our planet. Interaction of the geomagnetic field with the solar-wind and with matter of the Earth dissipates energy that is continuously re-supplied.

Generally, moving electric charges produce a magnetic field. For seventy years the geomagnetic field has been thought to be generated within the Earth's fluid core by a convection-driven dynamo (magnetic amplifier) mechanism (Elsasser 1939). Although I concur with the mechanism, I dispute its location. Long-term, stable convection in the Earth's fluid core would require rapid removal of heat brought to the top of the core by convection, which is impossible because the core is wrapped in an insulating rock-blanket 2900km thick (Appendix II).

My georeactor provides both the energy source and the convection-stable environment necessary for geomagnetic field generation (Herndon 1996, 2007, 2009). It provides a nuclear-fission heat source, the georeactor's uranium sub-core, surrounded by its fluid sub-shell composed radioactive waste (Figure 3). Heat from the nuclear sub-core causes convection in the electrically-conducting, fluid sub-shell. Planetary rotation twists the convective motions producing a dynamo, a magnetic amplifier, that amplifies a tiny "seed" magnetic field, produced by radioactive decay products, to generate the geomagnetic field. Heat brought to the top of the fluid sub-shell is removed by the massive heat sinks, the inner core and fluid core, thus assuring stable convection.

Solar System influences, including electrical currents induced by super-intense solar outbursts, I suggest, intermittently disrupt the georeactor's magnetic field generation. The georeactor (Figure 3), is one ten-millionth as massive as the fluid core so that disrupted convection leads to rapid changes, including reversals in the geomagnetic field. The occurrence of episodes of very rapid geomagnetic field change have been reported, six degrees per day during one reversal (Coe & Prevot 1989) and another of one degree per week (Bogue 2010).

## Summary

Evidence that the interior of Earth resembles an enstatite chondrite (and not an ordinary chondrite as long believed), began for me a logical progression of understanding that led, step-by-step, to a new vision of Earth's internal composition, early formation as a Jovian gas-giant, geodynamics with new, powerful energy sources, and nuclear georeactor magnetic field generation: In short, a new indivisible geoscience paradigm. My testable concept, anchored to well-established physical and chemical laws, to myriad astronomical and geological observations, to the cosmic abundance of the elements and to the properties of matter, leads to this whole-Earth decompression dynamics view that simultaneously solves several scientific problems. Clearly, critical observational, experimental, and theoretical evaluation is warranted.

## Acknowledgement

I thank Professor Lynn Margulis for editorial excellence, persistence, and insistence that all should be able to understand. I thank the institutions that made possible her generous expenditure of time and effort: STIAS (Stellenbosch Institute of Advanced Study, South Africa) and the University of Massachusetts-Amherst Graduate School, College of Natural Sciences, and I thank advanced students: Kendra Clark, James MacAllister, and Bruce Scofield. I thank Dr. Dennis Edgerley for graphic assistance.

## Appendix I: Physical Impossibility of Earth-Mantle Convection

Since the 1930's, convection has been assumed to occur within the Earth's mantle (Holmes 1931) and, since the 1960's has been incorporated as an absolutely crucial component of seafloor spreading in plate tectonics theory. Instead of looking questioningly at the process of convection, many have assumed without corroborating evidence that mantle convection "must" exist.

Chandrasekhar (1957) described convection in the following way: "The simplest example of thermally induced convection arises when a horizontal layer of fluid is heated from below and an adverse temperature gradient is maintained. The adjective 'adverse' is used to qualify the prevailing temperature gradient, since, on account of thermal expansion, the fluid at the bottom becomes lighter than the fluid at the top; and this is a top-heavy arrangement which is potentially unstable. Under these circumstances the fluid will try to redistribute itself to redress this weakness in its arrangement. This is how thermal convection originates: It represents the efforts of the fluid to restore to itself some degree of stability."

The lava lamp, invented by Smith (1968), affords an easy-to-understand demonstration of convection at the Earth's surface. Heat warms a blob of wax at the bottom, making it less dense than the surrounding fluid, so the blob floats to the surface, where it loses heat, becomes denser than the surrounding fluid and sinks to the bottom. Convection is applicable in circumstances wherein density is constant except as altered by thermal expansion; in the lava-lamp, for example, but not in the Earth's mantle. The Earth's mantle is "bottom-heavy", i.e., its density at the bottom is about 62% greater than its top (Dziewonski & Anderson 1981). The potential decrease in density by thermal expansion, <1%, cannot make the mantle "top-heavy" as described by Chandrasekhar. Thus mantle convection cannot be expected to occur.

Mantle convection is often (wrongly) asserted to exist on the basis of a high calculated Rayleigh Number (Rayleigh 1916), which was derived to quantify the onset of instability in a thin, horizontal layer of *incompressible* fluid of *uniform density*, except as altered by thermal expansion when heated from beneath. The Rayleigh Number is not applicable to the Earth's mantle, which is *neither incompressible nor of uniform density*.

It is instructive to apply the principle upon which submarines operate, "neutral buoyancy", to the Earth's mantle. The idea is that a heated "parcel" of bottom mantle matter, under the physically-unrealistic assumption of ideal, optimum conditions, will float upward to come to rest at its "neutral buoyancy", the point at which its own density is the same as the prevailing mantle density.

Consider a "parcel" of matter at the base of the Earth's lower mantle existing at the prevailing temperature, $T_0$, and having density, $\rho_0$ (Dziewonski & Anderson 1981). Now, suppose that the "parcel" of bottom-mantle matter is selectively heated to temperature $\Delta T$ degrees above $T_0$. The "parcel" will expand to a new density, $\rho_z$, given by

$$\rho_z = \rho_0 (1-\alpha\Delta T)$$

where α is the volume coefficient of thermal expansion at the prevailing temperature and pressure.

Now, consider the resulting dynamics of the newly expanded "parcel". Under the assumption of ideal, optimum conditions, the "parcel" will suffer no heat loss and will encounter no resistance as it floats upward to come to rest at its "neutral buoyancy", the point at which its own density is the same as the prevailing mantle density. Even at an extreme $\Delta T = 600°K$ the maximum float distance to neutral buoyancy is <25 km, just a tiny portion of the 2230 km distance required for lower mantle convection, and nearly 2900 km required for whole-mantle convection.

Decades of belief that mantle convection "must" exist has resulted in a plethora of mantle convection models that, of course, purport to show that mantle convection is possible under certain assumed conditions. Generally, models begin with a preconceived result that is invariably achieved through result-selected assumptions. Although rarely, if ever, stated explicitly, in convection models, the mantle is tacitly assumed to behave as an ideal gas.

Stellar convection models involved a gravitationally compressed system of $H_2$ and He gas at ~5000K that is thought to approach ideal gas behavior, i.e., no viscosity, hence, no viscous loss. In those models a heated parcel of ideal-gas expands and rises, never losing heat to its surroundings, and never coming to rest at "neutral buoyancy". The parcel maintains pressure equilibrium with its surroundings as it begins to rise, decompressing and expanding against progressively lower pressure, while maintaining its initial heat perturbation. The only impediment to such ideal-gas convection is if heat can be transported more rapidly by conduction and/or radiation than by convection.

Mantle convection models typically apply the same reasoning and assumptions as stellar convection models. A heated parcel of mantle matter is assumed to float ever-upward decreasing in density, never reaching "neutral buoyancy", while maintaining its heat content. But the mantle is not an ideal gas; it is a crystalline solid, not even a super-cooled liquid like glass. But, like its stellar counterpart, it assumed to behave "adiabatically", i.e., to maintain the parcel's initial heat perturbation, suffering no heat loss, even although in reality the mantle: (1) is extremely viscous and thus subject to viscous losses; (2) potentially moves by convection at a rate not too different from the rate heat is conducted; (3) has compositionally-different layers; (4) may have crystalline phase boundaries; and (5) possesses unknown rheological properties. Earthquakes, for example, occur within the mantle to depths of about 660km and signal the catastrophic release of pent-up stress. Processes and properties such as these, I submit, would readily block mantle convection.

## Appendix II: Physical Impossibility of Earth-Core Convection

Elsasser (1939) proposed that the geomagnetic field is generated by a convection-driven dynamo mechanism in the Earth's fluid core, and, since that time, Earth-core convection has been assumed. But Earth-core convection is physically impossible for the same reasons convection is physically impossible in the Earth's mantle (Appendix I), plus another reason related to heat transport.

Although the Earth's core is a fluid, it is "bottom-heavy", i.e., its density at the bottom is about 23% greater than at the top (Dziewonski & Anderson 1981). The potential decrease in density by thermal expansion, <1%, cannot make the fluid core "top-heavy", as described by Chandrasekhar (1957). Thus, Earth-core convection cannot be expected to occur. Moreover, the non-dimensional Rayleigh Number (Rayleigh 1916) is not applicable to the Earth's Core, which is *neither incompressible nor of uniform density*.

For sustained convection to occur, heat brought from the core-bottom must be efficiently removed from the core-top to maintain the "adverse temperature gradient" described by Chandrasekhar (1957), i.e., the bottom being hotter than the top. But, efficient heat removal is physically impossible because the Earth's core is wrapped in an insulating silicate blanket, the mantle, 2900 km thick that has significantly lower thermal conductivity, lower heat capacity, and greater viscosity than the Earth's core. Heat transport within the Earth's fluid core must therefore occur mainly by thermal conduction, not convection.

The geomagnetic implication is quite clear: Either the geomagnetic field is generated by a process other than the convection-driven dynamo-mechanism, or there exists another fluid region within the deep-interior of Earth which can sustain convection for extended periods of time. I have provided the reasonable basis to expect long-term stable convection in the georeactor sub-shell, and proposed that the geomagnetic field is generated therein by the convection-driven dynamo mechanism (Herndon 2007, 2009). Heat produced by the georeactor's nuclear sub-core causes convection in the surrounding fluid radioactive-waste sub-shell; heat is removed from the top of the sub-shell by a massive, thermally-conducting heat-sink (the inner core) that is surrounded by an even more massive, thermally-conducting heat-sink (the fluid core).

# References

Birch, F. 1940 The transformation of iron at high pressures, and the problem of the earth's magnetism. *Am. J. Sci.* **238**, 192-211.

Bogue, S. W. 2010 Very rapid geomagnetic field change recorded by the partial remagnetization of a lava flow *Geophys. Res. Lett.* **37**, doi: 10.1029/2010GL044286.

Cameron, A. G. W. 1963 Formation of the solar nebula. *Icarus* **1**, 339-342.

Carey, S. W. 1976 *The Expanding Earth*. Amsterdam: Elsevier.

Chandrasekhar, S. 1957 Thermal Convection. *Proc. Amer. Acad. Arts Sci.* **86**, 323-339.

Coe, R. S. & Prevot, M. 1989 Evidence suggesting extremely rapid field variation during a geomagnetic reversal. *Earth Planet. Sci. Lett.* **92**, 192-198.

Dalrymple, G. B. 1991 *The Age of the Earth*. Stanford: Stanford University Press.

Dziewonski, A. M. & Anderson, D. A. 1981 Preliminary reference Earth model. *Phys. Earth Planet. Inter.* **25**, 297-356.

Elsasser, W. M. 1939 On the origin of the Earth's magnetic field. *Phys. Rev.* **55**, 489-498.

Eucken, A. 1944 Physikalisch-chemische Betrachtungen ueber die frueheste Entwicklungsgeschichte der Erde. *Nachr. Akad. Wiss. Goettingen, Math.-Kl.*, 1-25.

Herndon, J. M. 1979 The nickel silicide inner core of the Earth. *Proc. R. Soc. Lond* **A368**, 495-500.

Herndon, J. M. 1993 Feasibility of a nuclear fission reactor at the center of the Earth as the energy source for the geomagnetic field. *J. Geomag. Geoelectr.* **45**, 423-437.

Herndon, J. M. 1994 Planetary and protostellar nuclear fission: Implications for planetary change, stellar ignition and dark matter. *Proc. R. Soc. Lond* **A455**, 453-461.

Herndon, J. M. 1996 Sub-structure of the inner core of the earth. *Proc. Nat. Acad. Sci. USA* **93**, 646-648.

Herndon, J. M. 2003 Nuclear georeactor origin of oceanic basalt $^3$He/$^4$He, evidence, and implications. *Proc. Nat. Acad. Sci. USA* **100**, 3047-3050.

Herndon, J. M. 2005 Whole-Earth decompression dynamics. *Curr. Sci.* **89**, 1937-1941.

Herndon, J. M. 2006a Energy for geodynamics: Mantle decompression thermal tsunami. *Curr. Sci.* **90**, 1605-1606.

Herndon, J. M. 2006b Solar System processes underlying planetary formation, geodynamics, and the georeactor. *Earth, Moon, and Planets* **99**, 53-99.

Herndon, J. M. 2007 Nuclear georeactor generation of the earth's geomagnetic field. *Curr. Sci.* **93**, 1485-1487.

Herndon, J. M. 2008 *Maverick's Earth and Universe*. Vancouver: Trafford Publishing.

Herndon, J. M. 2009 Nature of planetary matter and magnetic field generation in the solar system. *Curr. Sci.* **96**, 1033-1039.

Herndon, J. M. 2010 Impact of recent discoveries on petroleum and natural gas exploration: Emphasis on India. *Curr. Sci.* **98**, 772-779.

Herndon, J. M. & Suess, H. E. 1976 Can enstatite meteorites form from a nebula of solar composition? *Geochim. Cosmochim. Acta* **40**, 395-399.

Hilgenberg, O. C. 1933 *Vom wachsenden Erdball*. Berlin: Giessmann and Bartsch.

Hollenbach, D. F. & Herndon, J. M. 2001 Deep-earth reactor: nuclear fission, helium, and the geomagnetic field. *Proc. Nat. Acad. Sci. USA* **98**, 11085-11090.

Holmes, A. 1931 Radioactivity and Earth movements. *Trans. geol. Soc. Glasgow 1928-1929* **18**, 559-606.

Keil, K. 1968 Mineralogical and chemical relationships among enstatite chondrites. *J. Geophys. Res.* **73**, 6945-6976.

Kennet, B. L. N., Engdahl, E. R. & Buland, R. 1995 Constraints on seismic velocities in the earth from travel times *Geophys. J. Int.* **122**, 108-124.

Lehmann, I. 1936 P'. *Publ. Int. Geod. Geophys. Union, Assoc. Seismol., Ser. A, Trav. Sci.* **14**, 87-115.

Oldham, R. D. 1906 The constitution of the interior of the earth as revealed by earthquakes. *Q. T. Geol. Soc. Lond.* **62**, 456-476.

Ramdohr, P. 1964 Einiges ueber Opakerze im Achondriten und Enstatitachondriten. *Abh. D. Akad. Wiss. Ber., Kl. Chem., Geol., Biol.* **5**, 1-20.

Rayleigh, L. 1916 On convection currents in a horizontal layer of fluid, when the higher temperature is on the under side. *Phil. Mag.* **32**, 529-546.

Reed, S. J. B. 1968 Perryite in the Kota-Kota and South Oman enstatite chondrites. *Mineral Mag.* **36**, 850-854.

Ringwood, A. E. 1961 Silicon in the metal of enstatite chondrites and some geochemical implications. *Geochim. Cosmochim. Acta* **25**, 1-13.

Scheidegger, A. E. 1982 *Principles of Geodynamics*. Heidelberg: Springer-Verlag.

Seager, S. & Deming, D. 2010 Exoplanet Atmospheres. *Ann. Rev. Astron. Astrophys.* **48**, 631-672.

Smith, D. G. 1968 Display Devices. U. S. Patent 3,387,396.

Wegener, A. L. 1912 Die Entstehung der Kontinente. *Geol. Rundschau* **3**, 276-292.

Wiechert, E. 1897 Ueber die Massenverteilung im Inneren der Erde. *Nachr. K. Ges. Wiss. Goettingen, Math.-Kl.*, 221-243.